# Study of narrowband single photon emitters in polycrystalline diamond films

*Russell G. Sandstrom, Olga Shimoni, Aiden A. Martin and Igor Aharonovich\**

*School of Physics and Advanced Materials, University of Technology, Sydney, P.O. Box 123, Broadway, New South Wales 2007, Australia*

*\* corresponding author: igor.aharonovich@uts.edu.au*

**Abstract**
Quantum information processing and integrated nanophotonics require robust generation of single photon emitters on demand. In this work we demonstrate that diamond films grown by microwave plasma chemical vapour deposition on a silicon substrate host bright, narrowband single photon emitters in the visible – near infra-red spectral range. The emitters possess fast lifetime (~ several ns), absolute photostability, and exhibit full polarization at excitation and emission. Pulsed and continuous laser excitations confirm their quantum behaviour at room temperature, while low temperature spectroscopy is done to investigate their inhomogeneous broadening. Our results advance the knowledge of solid state single photon sources and open pathways for their practical implementation in quantum communication and quantum information processing.

The urgent demand to translate quantum information science into practical quantum technologies and devices accelerated the quest for an ideal solid state single photon source[1,2]. In addition to the traditional candidates such as quantum dots[3,4], defects in wide band-gap materials – including diamond[5], silicon carbide[6] and ZnO[7,8] have recently attracted considerable attention as potential single photon emitters (SPEs). The main advantage of these emitters is their photostability and room temperature operation.

In addition, equally important prerequisites of SPEs, are their fast emission lifetime and narrow full width at half maximum (FWHM), ideally Fourier transform limited. These photo-physical properties enable efficient filtering and optimization of signal to noise ratio that are paramount for practical devices. Some defects that meet these criteria include the chromium related[9] or the silicon vacancy (SiV)[10] colour centres in diamond. Both exhibit narrowband emission at the near infra-red (NIR), low phonon coupling and relatively fast emission lifetime (shorter than 5 ns). Several recent works have also identified other native NIR defects[11,12], however, robust, reproducible recipes to engineer these defects are still under development.

In this work we report on a previously unknown family of single photon emitters, that exhibit promising photophysical properties in the visible – near infrared range. Remarkably, these defects can be easily and cost effectively engineered in thin polycrystalline diamond films grown by microwave plasma chemical vapour deposition (MPCVD) on a silicon substrate. The emitters appear uniformly all across the film, in conjunction with the emission from an ensemble of SiV defects. Figure 1a shows a scanning electron microscope (SEM) image of the polycrystalline diamond film. The diamond film was grown on silicon substrates using MPCVD from 4 – 6 nm detonation nanodiamonds seeds[13]. The sample is homogeneous and the final grain size is smaller than 200 nm. The optical properties of the film were studied using a custom confocal microscope with Hanbury Brown and Twiss (HBT) interferometer to detect single photon emission. Figure 1b shows a typical room temperature photoluminescence (PL) spectrum recorded using a 532 nm excitation wavelength. A narrow band emission peak at ~ 685 nm, in addition to the typical SiV peak found at 738 nm are clearly visible. As will be shown, most of the observed narrow band emitters are SPEs. The origin of the silicon is the silicon substrate on which the diamond film is grown. During the growth, the silicon substrate is slightly etched, and the silicon atoms are incorporated into the film, forming the SiV defects.

We studied in more detail the emission properties from the diamond film. Multiple narrowband emitters were observed over a large range of wavelengths, from 560 nm to 800 nm. Figure 1(c, d) shows the statistical distribution of the narrow PL signals and the FWHM of the emission lines, respectively. The distribution of the zero phonon lines (ZPLs) varies, with some increase in occurrence at ~ 630 and 670 nm. In addition to the narrowband emitters, every spectrum exhibits the typical SiV peak at 738 nm, indicating that the signal originates from the diamond nanocrystals. Figure 2b shows the histogram of the FWHM of the narrowband emitters. Most of the investigated emitters have narrow bandwidth centred around ~ 3 nm, with negligible phonon sideband.

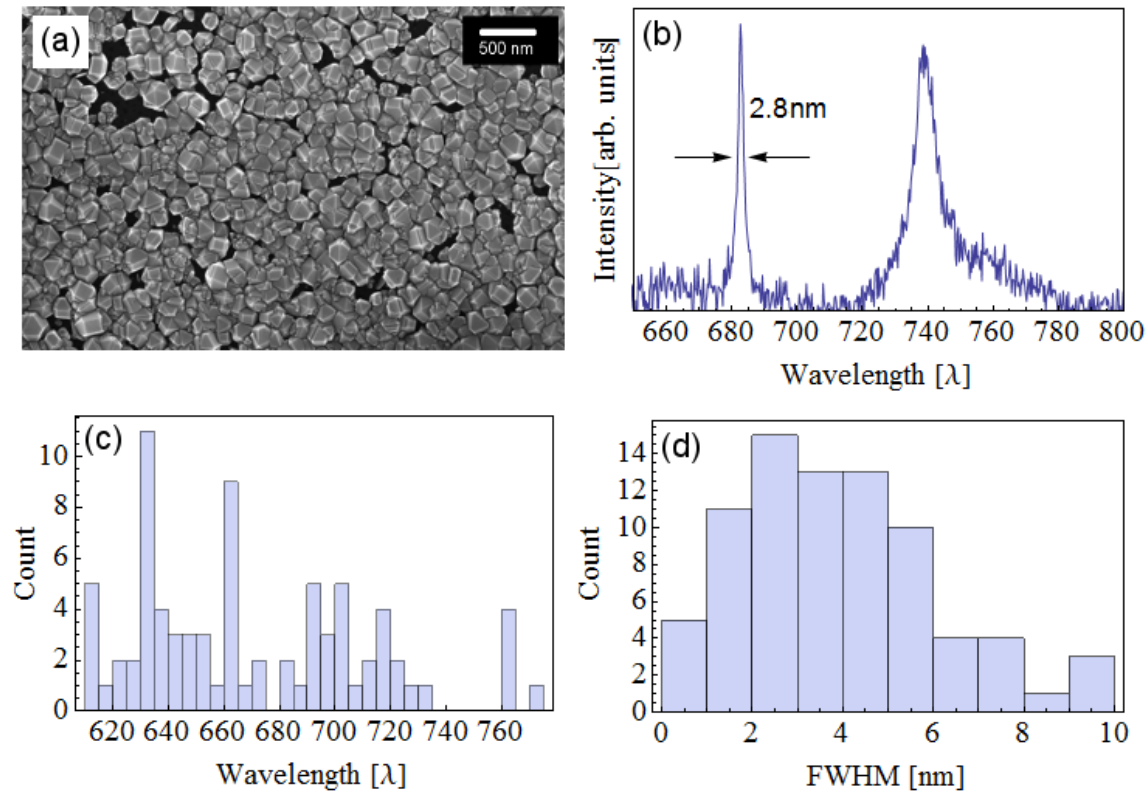

*Figure 1. (a) SEM image of the grown nanodiamonds film using MPCVD method. (b) A representative PL spectrum from the film, recorded at room temperature showing a narrowband emission at ~ 685 nm and the SiV emission at 738 nm. (c) Histogram of the peak positions of the narrowband emission lines. More than 100 emitters were evaluated. (d) Histogram showing the FWHM of the emitters. The emitters have predominantly narrowband emission (FWHM ~ 3 nm).*

To study the photoluminescence properties further, polarization measurements were undertaken. To carry out the polarization measurements, a Glen Taylor polarizer and a half wave plate (WPH05M-532, Thorlabs) were placed in the excitation pathway. The half wave plate was rotated to vary the excitation polarization of the incoming laser beam. Figure 2a displays an example of an excitation polarization measurement for an emitter with two emitting lines at 627 nm and at 684 nm. The absorbing transition of these emitters is likely to involve higher vibronic states in the excited state manifold, or higher excited electronic states of this particular defect[14]. As the SiV is always visible in addition to the studied narrow emitters, its involvement in the excitation pathways cannot be excluded. All the investigated emitters behave like a single dipole, with minima close to zero for excitation with linearly polarized light. The excitation polarization visibility is given by

$$V = \frac{I_{max} - I_{min}}{I_{max} + I_{min}}$$

where $I_{max}$, $I_{min}$ are the maximum and the minimum intensities, respectively. For this particular emitter, a visibility of 0.96 is measured. High visibility indicates that the absorbing dipole is aligned with the excitation wavelength dipole. However, since the investigated defects are located within a polycrystalline diamond film, it is expected that not all the emitters will show full polarization extinction, as was indeed observed in our experimental work. Since the nanodiamonds are oriented differently within the film, we cannot comment on the actual crystallographic orientation of the emitters within the films. We note that further deviations from 100% visibility are attributed to background luminescence of the diamond material at the vicinity of the defects and at the same wavelengths.

For the emission polarization, a linear polarizer (LPVIS050, Thorlabs) was inserted at the collection path. Figure 2 (b) displays the emission polarization of the same emitter, obtained by rotating only the polarizer at the collection path. Using the same equation, the emission visibility is calculated to be 0.86. We note that for the emission case, an interesting pattern evolves, where the emitter at 627 nm has parallel excitation and emission dipoles (within the error of our measurement), while the emitter at 684 nm undergoes a depolarization process. In the latter case, reorientation process in the excited state manifold occurs, and the final emission pattern, although fully polarized, is perpendicular to the excitation dipole. This behaviour was observed previously for several emitters in diamond[9, 15].

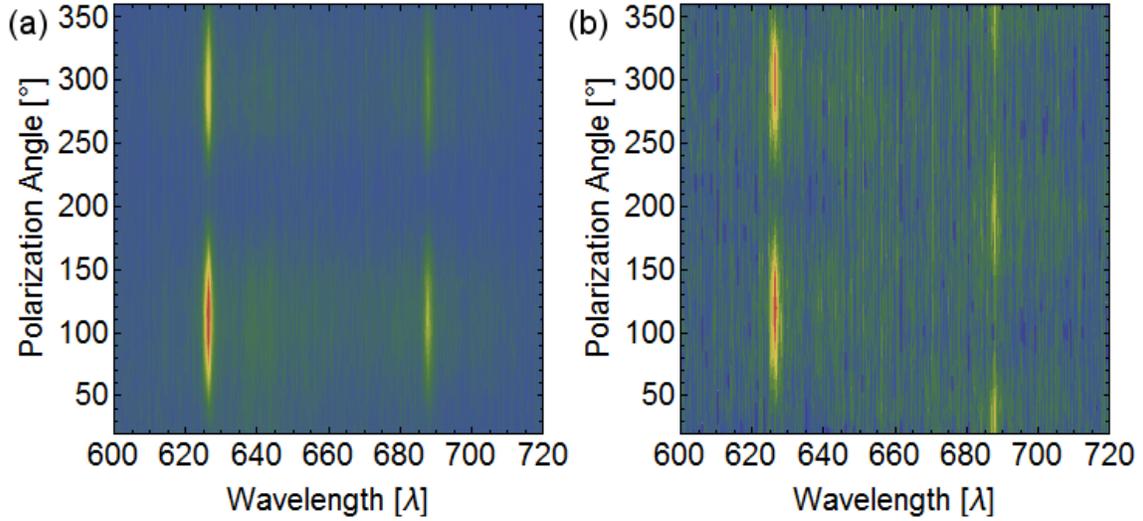

*Figure 2 (a) Excitation polarization and (b) Emission polarization of the narrowband emitters. The excitation polarization was measured by rotating a half wave plate at the excitation path, while the emission polarization was recorded by rotating a polarizer at the collection path.*

To prove that the narrow lines are quantum emitters, second order correlation measurements were performed using an HBT setup. Figure 3 (a) shows the normalised second order correlation function, $g^{(2)}(\tau)$, measured from a narrow line at 685 nm under continuous wave (CW) excitation at room temperature. To maximize the signal to noise, a 10 nm bandwidth filter (Semrock) was added at the collection path. The data was fit to a three level system using the equation:

where   and   are the excited state and the shelving state lifetimes of the emitter, respectively. The observed dip is below 0.5, which indicates that the emission originates from a single photon emitter. The slight bunching at shorter delay times, where   , indicates that a metastable (shelving) state exists[10, 14].

One of the key criteria for utilizing quantum sources for practical quantum computation and communication devices, is triggered excitation[4, 16, 17]. Figure 3 (b) shows a second order correlation function under pulsed excitation. For this measurement, a 514 nm picosecond laser with 40 MHz repetition rate was used. The dip at zero delay time,    , is below 0.5, corresponding to a triggered single photon emission. The deviation from 0 in both CW and pulsed excitation cases is attributed to the background florescence from the diamond film. With an increased excitation power to saturate the emitter, more than a million counts per second were recorded from this defect. This was achieved without use of external light enhancement devices, such as nanowires[18] or solid immersion lenses[19]. Similar count rates were achieved from other narrowband emitters found in the polycrystalline film. Finally, the emitters exhibited absolute photostability, and no blinking or bleaching were observed under our experimental conditions.

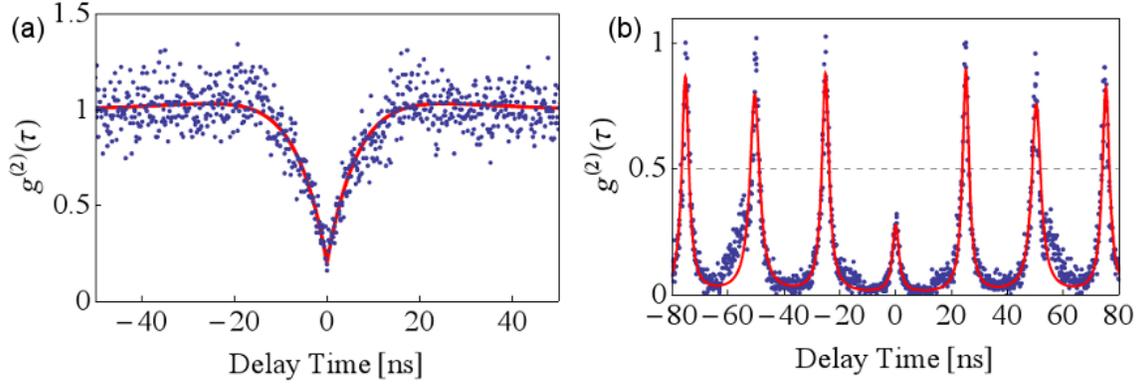

*Figure 3.* *Second order correlation function measured under (a) continuous wave optical excitation and (b) pulsed (triggered) optical excitation. The dip at zero delay time confirms the emitter is a single photon emitter. The deviation from a $g^{(2)}(0)=0$ is due to background luminescence from the diamond film. Blue dots are the experimental data and the red line is the fit.*

Next, we performed low temperature spectroscopy on the narrowband emitters. Figure 4(a) shows an example of a single narrowband emitter with a ZPL centred at 728 and a FWHM of 0.6 nm. Emission from an ensemble of SiV defects is shown for comparison (ZPL at 738 nm). Note that a single emitter is brighter than the ensemble of SiV defects. We probed numerous emitters, and the histogram of the FWHM of the emitters is shown as an inset in figure 4(b). Majority of the emitters have narrow emission lines, limited by our spectrometer (~ 0.2 nm).

The temperature dependence of the FWHM of a single emitter is shown in figure 4 (b). The emitters exhibit temperature dependent FWHM, that shrinks to sub nm at low temperature (0.3 nm corresponds to 185 GHz at 700 nm). The temperature dependent line width shift is an important parameter to understand the centre dynamics and its interaction with the crystal vibration[20-23]. For other defects in diamond, including the SiV and nitrogen vacancy (NV) centres, various temperature dependent mechanisms have been reported. The dynamic Jahn-Teller dependence ($T^5$) is predominant for the NV defect[24] and ($T^3$) dependence was reported for the SiV defect[23]. Our measurements indicate that $T^3$ dependence describes our results appropriately as can be seen in figure 4(b). This indicates that a significant inhomogeneous broadening of defects occurs through interaction with external impurities. Such an observation is plausible, since the polycrystalline film hosts many other defects, including nitrogen and silicon.

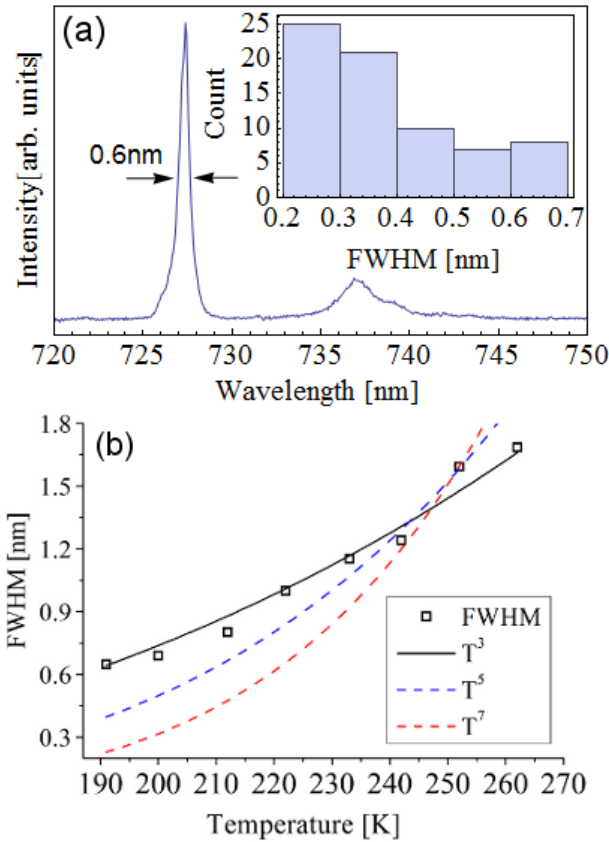

Figure 4. (a) Low temperature (5 K) PL spectrum recorded from the diamond film, showing the narrowband, bright emission from a single defect. Inset is the histogram of FWHM values of the narrowband defects, measured at low temperature of 5 K. (c) The temperature dependent FWHM values (black squares) as a function of temperature from 180 K to 270 K. The best fit is observed for $T^3$ (black line), while fits of $T^5$ and $T^7$ are shown for comparison as dotted blue and red lines, respectively.

Finally, we comment on the potential origin of these SPEs. The emitters were observed in polycrystalline diamond films, grown using a conventional MPCVD technique with no annealing steps after the growth and in various films grown under radically different growth conditions in different reactors. The parameters for the growths were (a) microwave power 3000 W, pressure 150 torr, $H_2/CH_4$ flow 10/750 standard cubic centimetres per minute (sccm) and (b) microwave power 900 W, pressure 60 torr, $H_2/CH_4$ flow 4/400 sccm. There have been previous reports on unidentified narrowband emitting defects at the NIR employed single crystal diamond samples or ultra-pure materials[11, 12]. The defects were excited using a red excitation wavelength, no SiV emission was observed in conjunction with the emitters, and an annealing step was required to activate these defects. These three factors lead to a conclusion that the single emitters reported in the literature were correlated with some impurities in the diamond – including residuals of transition metals. In our case, however, green excitation was used to excite and probe the narrowband defects. Furthermore, we have clearly shown that these defects are formed in conjunction with the SiV defect, in standard, routine polycrystalline diamond films. This observation, along with the fact that the emitters were reproducibly observed under different growth conditions, indicates that the SPEs are not likely to be associated with transition metals.

As a control experiment, we studied isolated diamond nanocrystals that were grown under identical conditions. To achieve isolated crystals, the seeding density was reduced. We do not observe narrowband defects in the nanocrystals, the only fluorescence feature present was the emission from the SiV defects. We therefore conclude that having a grain boundary enhances the probability of observing these defects. In this regard hydrogen may play a crucial role, as it is known that hydrogen defects in diamond are located at the grain boundaries[25] and an interaction between point defects and grain boundaries has also been reported for diamond[26]. Finally, under our experimental growth conditions, nitrogen can be easily incorporated into the film. While the nitrogen is not necessarily involved in the emission pathway, and the emission from the NV center is not observed, it may act as a donor or stabilize the local environment and change the Fermi level within the nanodiamonds, therefore opening pathways to generation of quantum emission from a localized state.

To summarize, we discovered a new family of narrowband, fully polarized single photon emitters that can be easily found in a standard polycrystalline diamond film. Not only do these emitters exhibit outstanding photophysical properties that can be harnessed to realize indistinguishable photons[27, 28], their abundance in low cost polycrystalline diamond films opens up opportunities to exploit quantum technologies without the need of rigours processing and ultrapure samples. Our work contributes towards a broader understanding of single photon sources in diamond and other wide band gap materials.

The authors thank Milos Toth for useful discussions. Dr Aharonovich is the recipient of an Australian research council discovery early career research award (project number DE130100592. Aiden A. Martin is the recipient of a John Stocker Postgraduate Scholar ship from the Science and Industry Endowment Fund.